\documentclass[useAMS,usenatbib,onecolumn,usegraphicx]{mn2e}
\usepackage{times}
\usepackage{url}
\usepackage{color}
\newcommand{\om}{\Omega_\rmn{m}}
\newcommand{\omh}{\om h^2}
\newcommand{\ode}{\Omega_\rmn{de}}
\newcommand{\ob}{\Omega_\rmn{b}}
\newcommand{\obh}{\ob h^2}

\newcommand{\bl}{\bmath{l}}
\newcommand{\bk}{\bmath{k}}
\newcommand{\simgt}{\lower.5ex\hbox{$\; \buildrel > \over \sim \;$}}
\newcommand{\simlt}{\lower.5ex\hbox{$\; \buildrel < \over \sim \;$}}

\begin{document}
\title[Cosmology with WL power spectrum and bispectrum tomography]
{
Cosmological parameters from weak lensing power spectrum and bispectrum
tomography: including the non-Gaussian errors
}

\author[I. Kayo and M. Takada]
{ 
Issha Kayo$^{1}$\thanks{E-mail: kayo@ph.sci.toho-u.ac.jp}
and Masahiro Takada$^{2}$\thanks{E-mail: masahiro.takada@ipmu.jp}   \\
$^1$ Department of Physics, Toho University, 2-2-1 Miyama, Funabashi, 
Chiba 274-8510, Japan
\\
$^2$ Kavli Institute for the Physics and Mathematics of the Universe
 (Kavli IPMU, WPI),  The University of Tokyo, Chiba 277-8582, Japan
} 

\maketitle

\begin{abstract}
 We re-examine a genuine power of weak lensing bispectrum tomography for
 constraining cosmological parameters,
 when combined with the  power spectrum tomography, based on the
 Fisher information matrix formalism.
To account for the full information at two- and three-point levels, we
 include all the power spectrum and bispectrum information built from
 all-available combinations of tomographic redshift bins, 
 multipole bins and different triangle configurations over a range of
 angular scales (up to $l_{\rm max}=2000$ as our fiducial choice).  For
 the parameter forecast,
we use the halo model approach in \citet*{Kayoetal:13} to model
 the non-Gaussian error covariances as
 well as the cross-covariance between the power spectrum and the
 bispectrum, including the halo sample variance or the nonlinear version
 of beat-coupling.
 We find that
 adding the bispectrum information leads to about 60\% improvement in
 the dark energy figure-of-merit compared to the lensing power spectrum
 tomography alone, for three redshift-bin tomography and a Subaru-type
 survey probing galaxies at typical redshift of $z_s\simeq 1$. The
 improvement is equivalent to a 1.6 larger survey area.  Thus our
 results show that the bispectrum or more generally any three-point
 correlation based statistics carries complementary information on
 cosmological parameters to the power spectrum. However, the improvement
 is modest compared to the previous claim derived using the Gaussian
 error assumption, and therefore our results imply less additional information in
 even higher-order moments such as the four-point correlation function.
\end{abstract}

\section{Introduction}

Cosmic acceleration is perhaps the most tantalizing problem in
cosmology.  Within Einstein's gravity theory, general relativity, the
observed cosmic acceleration can be explained by introducing dark
energy, which acts as a repulsive force to accelerate the cosmic
expansion.  Alternatively, it might be a signature of the breakdown of
general relativity on cosmological scales \citep[see][for a
review]{JainKhoury:10}. Many on-going and upcoming wide-area galaxy
surveys aim at testing dark energy and modified gravity scenarios as the
origin of cosmic acceleration \citep[see][for a
review]{Weinbergetal:12}.  These range from ground-based imaging surveys
such as
the Panoramic Survey Telescope \& Rapid Response System
(Pan-STARRS\footnote{\url{http://pan-starrs.ifa.hawaii.edu}}), the Very Large Telescope Survey Telescope (VST)
Kilo-Degree
Survey (KiDS)\footnote{\url{http://www.astro-wise.org/projects/KIDS/}}, the
Subaru Hyper Suprime-Cam (HSC) Survey
\citep{Miyazakietal:12}\footnote{\url{http://www.naoj.org/Projects/HSC/index.html}},
the Dark Energy Survey
(DES\footnote{\url{http://www.darkenergysurvey.org}}), and 
 the Large Synoptic Survey Telescope (LSST\footnote{\url{http://www.lsst.org}}) to space-based missions such
as the European Space Agency (ESA) Euclid
mission\footnote{\url{http://sci.esa.int/science-e/www/area/index.cfm?fareaid=102}}
and National Aeronautics and Space Administration (NASA) Wide-Field Infrared Survey Telescope (WFIRST) satellite mission \citep{Spergeletal:13}
\footnote{\url{http://wfirst.gsfc.nasa.gov/}}.

Weak gravitational lensing or
cosmic shear is recognized as one of the most promising methods for
constraining cosmology
\citep[see][for
reviews]{BartelmannSchneider:01,SchneiderBook:06,HoekstraJain:08}. 
Since weak lensing directly probes the total matter distribution in
the large-scale structure, free
of galaxy bias uncertainty, it allows for a relatively clean
comparison of the measurement with theory.
The cosmological constraints based on the weak lensing measurements have
been reported by 
several groups \citep{Hamanaetal:03,Schrabbacketal:10,
HoekstraJain:08} and more recently by the Canada--France--Hawaii Telescope (CFHT) Lens Survey
\citep{Kilbingeretal:13,Heymansetal:13} 
and the
Planck collaboration \citep{PlanckWL:13}. 

However, the useful cosmological information in the weak lensing field
is mainly from the nonlinear clustering regime, over the range of
multipoles around $l\simeq$ a few thousands
\citep{JainSeljak:97,HutererTakada:05}.  Due to mode-coupling nature of
the nonlinear structure formation, the weak lensing field at angular scales
of interest displays non-Gaussian features. Hence the two-point
correlation function or its Fourier counterpart, power
spectrum, can no longer carry the full information of the weak lensing
field, unlike in the cosmic microwave background (CMB). Using
ray-tracing simulations and/or analytical methods such as the halo model
approach, previous work has shown that the non-Gaussianity causes
significant correlations between the power spectrum amplitudes at
different multipoles
\citep{WhiteHu:00,CoorayHu:01,Sembolonietal:07,Satoetal:09,Satoetal:10,TakadaJain:09,HarnoisDerapsetal:12,Kayoetal:13}. In
particular, \cite{Satoetal:09} used 1000 ray-tracing simulation
realizations to directly compute the power spectrum covariance for a
$\Lambda$-dominated cold dark matter ($\Lambda$CDM) 
model, and then showed that, for a
survey probing galaxies at typical redshift of $z_s\simeq1 $, the
non-Gaussian error covariance degrades the information content of weak
lensing power spectrum by a factor of 2--3 up to the maximum multipole
of a few thousands compared to the Gaussian information of the initial
density field. It was shown that a significant contribution of the
non-Gaussian errors arises from {\em the halo sample variance} (HSV) due
to super-survey modes of length scales comparable with or larger than a
survey size, which is an unobservable mode \citep[see
also][]{Hamiltonetal:06,TakadaBridle:07,TakadaJain:09,Takahashietal:09,Kayoetal:13,TakadaHu:13}. A
physical interpretation of the HSV effect is as follows.  If a survey
region is embedded in a coherent over- or under-density region, the
abundance of massive halos is up- or down-scattered from the
ensemble-averaged expectation according to halo bias theory or the
peak-background split theory
\citep{MoWhite:96,Moetal:97,ShethTormen:99,HuKravtsov:03}. Then the
modulation of halo abundance causes up- or down-scatters in the
amplitudes of weak lensing power spectrum at the small scales.

How can we recover the information content of the weak lensing field
beyond the power spectrum? Is the the initial Gaussian information lost at
small scales due to the highly nonlinear mode-coupling? Clearly some
of the initial Gaussian information should be encoded in higher-order
correlation functions of the weak lensing field, which carry
complementary information that cannot be extracted by the power spectrum
\citep{TakadaJain:03,TakadaJain:03a,Sembolonietal:11,TakadaJain:04,Kayoetal:13,SatoNishimichi:13}. The
three-point correlation function or its Fourier counterpart, the
bispectrum, is the lowest-order correlation that can extract the
non-Gaussian information. In addition, since the bispectrum or more
generally the three-point correlation based statistics depends on
cosmological parameters in a different way from the power spectrum,
adding the bispectrum information help to lift parameter degeneracies
\citep{Bernardeauetal:97,JainSeljak:97,Hui:99,Jainetal:00,WhiteHu:00,
HamanaMellier:01,VanWaerbekeetal:01,CoorayHu:01b,TakadaJain:02,
TakadaJain:04, DodelsonZhang:05,KilbingerSchneider:05,
Sembolonietal:08,Bergetal:10,Munshietal:11,Piresetal:12}. The first
attempt to measure the non-Gaussian signals from actual data was made by
several groups \citep{BernardeauMellieretal:02,Zhangetal:03,Jarvisetal:04}.
\citet{Sembolonietal:11} recently reported a detection of the skewness
from the Cosmological Evolution Survey (COSMOS) data, and showed an improvement in cosmological
parameters when combined with the two-point correlation constraints.

However, to realize the genuine power of the weak lensing bispectrum,
we
need to include all the lensing bispectra of different triangle
configurations available over a range of angular scales. Further, when
adding tomographic redshift information -- the so-called lensing
tomography \citep{Hu:99,Huterer:02,TakadaJain:04}, we need to 
include the bispectra built from different combinations of redshift bins
for each triangle configuration. Thus the number of different bispectra
can easily go beyond $10^3$ or $10^4$ (we will consider up to nearly $10^4$
bispectra in this paper). In order to properly count the independent
information of the power spectrum and bispectrum and not to double-count
their information, we need to compute the covariance matrices including
the HSV effect. If we want to use ray-tracing simulations to compute the
covariance matrix for all the bispectra, it requires a huge number of
the simulation realizations for each cosmological model, which is still
challenging \citep[see][for the first attempt for a reduced number of
bispectra]{SatoNishimichi:13}.
In our previous paper \citep{Kayoetal:13}, we developed the analytical
method to model the bispectrum covariance, based on the halo model
approach, and then showed that the model predictions fairly well
reproduce the covariance measured from the 1000 simulation realizations,
yet without tomography (we worked on 204 bispectra). 
It was shown that the bispectrum adds the
information content to the power spectrum, but the combined measurement
does not fully recover the Gaussian information mostly due to the HSV
contamination, i.e. super-survey modes.

The purpose of the paper is to extend the method in \citet{Kayoetal:13}
to lensing tomography case and to estimate an ability of upcoming
lensing surveys for constraining cosmological parameters with the
lensing power spectrum and bispectrum tomography. To do this, we employ
the halo model based method to properly account for the non-Gaussian
error covariances and include all the two- and three-point level
information, i.e. all the power spectrum and bispectra constructed from
different combinations of multipole bins, redshift bins and triangle
configurations. Hence, this work can be considered as a comprehensive
revisit of \citet{TakadaJain:04}, where the Gaussian error covariance
was assumed in the parameter forecast calculation.

The structure of this paper is as follows. In Section~\ref{sec:powerbi},
we develop the analytical model to describe the power spectrum and
bispectrum covariances and their cross-covariance when including lensing
tomography information, based on the halo model. Then we also describe
the Fisher information matrix formalism, which we use to estimate an
ability of future surveys for constraining cosmological parameters with
the lensing observables. In Section~\ref{sec:results} we show the
parameter forecasts.
 Section~\ref{sec:conclusion} is devoted to conclusion and
discussion.

\section{Lensing power spectrum and bispectrum tomography}
\label{sec:powerbi}

\subsection{Lensing power spectrum and bispectrum}
\label{sec:tomography}

Suppose that $\kappa_{(i)}(\bmath{\theta})$ is the lensing
convergence field at an angular position $\bmath{\theta}$ on the sky,
which is measurable from statistical distortion of source galaxies
residing in the
$i$-th tomographic redshift bin.
The convergence field is obtained by 
a weighted projection of the three-dimensional matter density
fluctuation field between the source galaxies and an observer
\citep[see][for a review]{BartelmannSchneider:01}:
\begin{equation}
 \kappa_{(i)}(\bmath{\theta})=\int_0^{\chi_H}\rmn{d}\chi
  W_{(i)}(\chi)\delta_m[\chi,\chi\bmath{\theta}],
\end{equation}
where $\chi$ is the comoving distance, $\chi_H$ is that to
the Hubble horizon and $\delta_m[\chi,\chi\bmath{\theta}]$ is the
three-dimensional matter 
fluctuation field. 
In the weak lensing regime,
the convergence field is equivalent to the
lensing shear field, which is given by the tidal field of large-scale
structure. 
The lensing efficiency function
$W_{(i)}(\chi)$ is given as
\begin{equation}
 W_{(i)}(\chi) = \frac{3}{2}\Omega_{\rm
 M}H_0^2a^{-1}(\chi)\chi
 \frac{1}{\bar{n}_{(i)}}\int_\chi^{\chi_H}\!\!\rmn{d}
  \chi_s n_{(i)}(z)\frac{\rmn{d} z}{\rmn{d} \chi_s}\frac{\chi_s-\chi}{\chi_s},
\end{equation}
where $n_{(i)}(z)$ is the redshift distribution of source galaxies in
the $i$-th tomography bin. In this paper, we simply employ a top-hat like
division of the galaxy distribution for lensing tomography; 
$n_{(i)}(z)$ is non-zero if $z$ resides in the $i$-th redshift
bin, 
$z\in [z_{i,{\rm lower}},z_{i,{\rm upper}}]$, otherwise $n_{(i)}(z)=0$.
The mean density of the source galaxies per unit solid angle,
$\bar{n}_{(i)}$, is given as 
\begin{equation}
 \bar{n}_{(i)}=\int_0^{\chi_H}\!\!\rmn{d} \chi_s n_{(i)}(z) \frac{\rmn{d} z}{\rmn{d} \chi_s},
\end{equation}
and this is used to model the shape noise contamination to the
error covariance matrices 
(see
Section~\ref{sec:cov}). 
For the whole redshift distribution of imaging galaxies, we simply
employ the following analytic form:
\begin{equation}
 n(z) \propto \frac{z^2}{2z_0^3}\exp\left(-\frac{z}{z_0}\right).
  \label{eq:nz}
\end{equation}
The parameter $z_0$
needs to be specified to resemble a hypothetical galaxy survey; the mean
redshift is given as $\langle z_s \rangle =3z_0$. We will assume 
$3z_0=1$ for a Subaru HSC-like survey, and $3z_0=0.7$ for a Euclid-lie survey,
respectively. 
For lensing tomography case, we denote $n_{(i)}(z)$ for the galaxy
distribution in the $i$-th redshift bin. 

Under the flat-sky approximation, 
the power spectrum and higher-order
correlation functions
of the convergence field are
defined in terms of the ensemble averages 
as
\begin{eqnarray}
 \langle \tilde\kappa_{(i)\bl_1} \tilde\kappa_{(j)\bl_2}\rangle
  &\equiv& (2\pi)^2 P_{(ij)}(l_1)\delta_\rmn{D}(\bl_1+\bl_2),\\
 \langle \tilde\kappa_{(i)\bl_1} \tilde\kappa_{(j)\bl_2} \tilde\kappa_{(k)\bl_3}\rangle
  &\equiv& (2\pi)^2 B_{(ijk)}(l_1, l_2, l_3)\delta_\rmn{D}(\bl_1+\bl_2+\bl_3),\\
 \langle \tilde\kappa_{(i_1)\bl_1} \tilde\kappa_{(i_2)\bl_2} \tilde\kappa_{(i_3)\bl_3}\tilde\kappa_{(i_4)\bl_4}\rangle_c
  &\equiv& (2\pi)^2 T_{(i_1i_2i_3i_4)}(\bl_1, \bl_2, \bl_3, \bl_4)\delta_\rmn{D}(\bl_1+\bl_2+\bl_3+\bl_4),\\
 \langle \tilde\kappa_{(i_1)\bl_1} \tilde\kappa_{(i_2)\bl_2} \cdots\tilde\kappa_{(i_n)\bl_n}\rangle_c
  &\equiv& (2\pi)^2 P_{n(i_1i_2\cdots i_n)}(\bl_1, \bl_2, \cdots,
  \bl_n)\delta_\rmn{D}(\bl_1+\bl_2+\cdots+\bl_n), \hspace{1cm}\mbox{ for } n\ge 5,
\end{eqnarray}
where $\tilde\kappa_{(i)\bl}$ is the Fourier-transformed coefficients of
the convergence field, defined as, $\tilde\kappa_{(i)\bl}=\int\!\!\rmn{d}^2\bmath{\theta}~
\kappa_{(i)}(\bmath{\theta})\exp(-\rmn{i}\bl\cdot\bmath{\theta})
$,
and $\delta_\rmn{D}(\bmath{k})$ is the two-dimensional Dirac delta function.  
 $P(l)$ is
the weak lensing power spectrum, $B(l_1,l_2,l_3)$ is the bispectrum and $P_n$ is the
$n$-point correlation function in Fourier space. 
The delta function $\delta_D(\bl_1+\bl_2+\cdots+\bl_n)$ in each equation
enforces the
condition that a set of $n$ vectors $(\bl_1,\bl_2,\cdots,\bl_n)$ forms
the closed $n$-point configuration in Fourier space. 
The ensemble
average denoted as $\langle\cdots\rangle_c$ is the connected part of the
higher-order correlation, the part which cannot be described by products
of the lower-order correlation functions \citep[e.g., see][for a
review]{Bernardeauetal:02}.
Due to statistical homogeneity and isotropy for the lensing field, the
power spectra obey the parallel translation symmetry (imposed by 
$\sum\bl_i=\bmath{0}$) 
as well as the
rotational symmetry of $n$-point configuration in Fourier space. 
Since each wavevector ($\bl_i$) has two degrees of
freedom in a two-dimensional case, the $n$-point correlation function is
specified by $2n-3$ parameters; $2n$ parameters for the $n$ wavevectors
minus 2 from the condition $\sum \bl_i=\bmath{0}$ and minus 1 for the
rotational symmetry. The symmetry constraints read that the power
spectrum (two-point correlation) is specified by 1 parameter, 
as $P_{(ij)}(l)$, where $l$ is the length of the wavevector, while 
the bispectrum (three-point correlation) is specified by 3 parameters,
e.g. the three side lengths of triangle configuration, ($l_1,l_2,l_3$).

These lensing spectra can be given as the weighted
line-of-sight projection of the three-dimensional spectra
of the underlying matter distribution. Using the Limber's
approximation \citep{Limber:54}, we can express
the $n$-point  power spectra of the weak lensing field
as
\begin{equation}
 P_{(ij)}(l)=\int_0^{\chi_H}\!\rmn{d}\chi
  W_{(i)}(\chi)W_{(j)}(\chi)\chi^{-2}P_\rmn{m}\!\left(k=\frac{l}{\chi};
					       \chi\right),
\end{equation}
\begin{equation}
 B_{(ijk)}(l_1, l_2, l_3)
  =\int_0^{\chi_H}\!\rmn{d}\chi
  W_{(i)}(\chi)W_{(j)}(\chi)W_{(k)}(\chi)\chi^{-4}
  B_\rmn{m}\!\left(k_1, k_2, k_3;
 \chi\right), \label{eq:bispectrum}
\end{equation}
and
\begin{equation}
 P_{n(i_1i_2\cdots i_n)}(\bl_1,\bl_2,\cdots,\bl_n)
  =\int_0^{\chi_H}\!\rmn{d}\chi
  W_{(i_1)}(\chi)W_{(i_2)}(\chi)
\cdots \times W_{(i_n)}(\chi)
\chi^{-2(n-1)}
P_{n}\!\left(\bk_1,\bk_2,\cdots,\bk_n;
 \chi\right), 
\end{equation}
where 
$k_i=l_i/\chi$, and $P_\rmn{m}$, $B_\rmn{m}$  and $P_n$
denote the  power spectrum,
bispectrum and $n$-point correlation function 
of the matter distribution at each redshift $\chi
(=\chi(z))$, respectively. 

When considering lensing tomography of $n_s$ redshift bins, we need to
account for different spectra 
for each multipole bin or $n$-point
configuration 
in
order to include the full information carried by the spectra. For the
power spectrum there are $n_s(n_s+1)/2$ spectra for 
each
multipole
bin $l$; e.g., for the case of two redshift bins, we need to include three
spectra, $P_{(11)}$, $P_{(12)}$ and $P_{(22)}$ for each $l$.
We need not consider $P_{(21)}$ 
because it is identical to
$P_{(21)}$.  
The bispectrum case is complicated. 
For a general triangle configuration with $l_1\neq l_2 \neq l_3$, 
we need to include $n_s^3$ bispectra for each triangle configuration of
$(l_1,l_2,l_3)$. For two redshift bin case ($n_s=2$), 
we have $B_{(111)}$, $B_{(112)}$, $B_{(121)}$, $B_{(122)}$, $B_{(211)}$,
$B_{(212)}$, $B_{(221)}$ and $B_{(222)}$\footnote{The bispectra are
different in a sense that their estimators are constructed from
different combinations of the Fourier coefficients such as 
$\tilde{\kappa}_{(i_1)\bl_1}
\tilde{\kappa}_{(i_2)\bl_2}
\tilde{\kappa}_{(i_3)\bl_3}
$ (see Eq.~15 in \cite{Kayoetal:13}). 
However, note that the ensemble-averaged expectation values are
identical: e.g., $\langle\hat{B}_{(i_1i_2i_3)}(l_1,l_2,l_3)\rangle = 
\langle\hat{B}_{(i_2i_1i_3)}(l_1,l_2,l_3)\rangle$ 
(see Eq.~\ref{eq:bispectrum}), but
their error covariances are
indeed different.}.
For an isosceles triangle
configuration such as $l_1=l_2(\neq l_3)$, the bispectrum estimators
constructed from 
$\tilde\kappa_{(i)\bl_1} \tilde\kappa_{(j)\bl_2}
\tilde\kappa_{(k)\bl_3}$ have symmetry
under permutation of
$\tilde\kappa_{(i)\bl_1} \leftrightarrow \tilde\kappa_{(j)\bl_2}$ or
$\{i,\bl_1\}\leftrightarrow\{j,\bl_2\}$, which reduces the number of
different bispectra for each set of $(l_1,l_2,l_3)$.
Thus each isosceles triangle configuration yields 
$n_s^2(n_s+1)/2$ bispectra. 
For 2 redshift bin case ($n_s=2$), 
there are six different bispectra;
$B_{(111)}$, $B_{(112)}$, $B_{(121)}$, $B_{(122)}$, $B_{(221)}$ and
$B_{(222)}$. 
For an equilateral triangle configuration, we need to consider 
$n_s(n_s+1)(n_s+2)/6$ for each triangle with side lengths $l_1=l_2=l_3$
due to further symmetries: 
$B_{(111)}$,
$B_{(112)}$, $B_{(122)}$ and $B_{(222)}$ for $n_s=2$. 
Thus, to take account of the full information carried by the 
bispectra for lensing tomography case, we need to include different
bispectra, but
need to avoid a double counting of identical bispectra, where we mean by
``identical'' that the ensemble averages of different bispectrum
estimators are identical 
and their covariance elements are also identical.

The power spectrum measurement for an actual survey is affected by 
intrinsic shape noise.
Assuming the Gaussian random shape noise (shapes of different galaxies are
uncorrelated with each other)
or equivalently ignoring the intrinsic
alignments in between different galaxies, 
the observed lensing power spectrum is contaminated by the
shape noise as
\begin{equation}
 P_{(ij)}^\rmn{obs}(l)=P_{(ij)}(l)
  +\delta^K_{ij}\frac{\sigma_\epsilon^2}{\bar{n}_{(i)}},
\label{eq:ps_obs}
\end{equation}
where $\sigma_\epsilon$ is the rms of intrinsic ellipticities per
component and $\delta^K_{ij}$ is the Kronecker delta function; 
$\delta_{ij}^K=1$ if $i=j$, otherwise $\delta^K_{ij}=0$.  
The Kronecker delta function enforces the condition that the shape noise
is present when considering correlations between the shapes of galaxies
in the same redshift bin, which thus arise from the same galaxy. In
other words, the shape noise is absent for cross-correlations of the
galaxy shapes in different redshift bins. 
The bispectrum and the higher-order spectra
are not affected by the shape noise, although their covariances have
the shape noise contamination.

\subsection{Error covariance matrix}
\label{sec:cov}

The covariance matrix describes a measurement accuracy of the lensing
spectrum for a given survey. We can extend the formulation of lensing
covariance matrices developed in \citet{Kayoetal:13} to the case of
lensing power spectrum and bispectrum tomography. 

\subsubsection{Power spectrum covariance}
The covariance matrix for the lensing power spectrum with tomographic
redshift bins is found to be
\begin{eqnarray}
\rmn{Cov}[P_{(ij)}(l),P_{(i'j')}(l')]
 &=&\rmn{Cov}_\rmn{Gauss}^\rmn{PS}+
 \rmn{Cov}_\rmn{NG}^\rmn{PS}+\rmn{Cov}_\rmn{HSV}^\rmn{PS} \nonumber \\
 &=&\frac{\delta^K_{ll'}}{N_\rmn{pairs}(l)}
 \left[ P_{(ii')}^\rmn{obs}(l)P_{(jj')}^\rmn{obs}(l)
 + P_{(ij')}^\rmn{obs}(l)P_{(ji')}^\rmn{obs}(l)\right]
 + \frac{1}{\Omega_\rmn{s}} \int\!\frac{\rmn{d} \psi}{2\upi}
 T_{(iji'j')}(\bl,-\bl,\bl',-\bl'; \psi)\nonumber\\
&&+\int\!\!\rmn{d}\chi W_{(i)}W_{(j)}W_{(i')}W_{(j')}\chi^{-4}
P_{\rmn{m}}^{\rm 1hb}(l/\chi;\chi)P_{\rmn{m}}^{\rm 1hb}(l'/\chi;\chi)
\int\!\!\frac{k\rmn{d}k}{2\upi}P_{\rmn{m}}^\rmn{L}(k;\chi)\left|
\tilde{W_\rmn{s}}(k\chi\Theta_{\rm s})
\right|^2,
\label{eq:pscovall}
\end{eqnarray}
where $\psi$ is the angle between the two vectors $\bl$ and $\bl'$ and
$\Omega_{\rm s}$ is the survey area. 
The
quantity $N_\rmn{pairs}(l)$ is the number of independent pairs of two
vectors $\bl$ and $-\bl$ in Fourier space, where the vector $\bl$ has
the length $l$ within the bin width $\Delta l$ and `independent' means
different pairs discriminated by the fundamental Fourier mode of a given
survey, $l_f\simeq 2\upi/\Theta_{\rm s}$ ($\Theta_{\rm s}$ is the
angular scale of the survey area).  At the limit $l_i\gg l_f$,
\begin{equation}
N_\rmn{pairs}(l)\simeq \frac{2\upi l \Delta l}{(2\upi/\Theta_{\rm s})^2}
=\frac{\Omega_{\rm s}l\Delta l
}{2\upi}.
\label{eq:n_mode}
\end{equation}
For the third term on the r.h.s., we have defined the notation
$P_{\rmn{m}}^{\rm 1hb}$ to denote the 1-halo term of  matter power
spectrum, weighted by the halo bias: 
\begin{equation}
P_{\rmn{m}}^{\rm 1hb}(k;\chi)\equiv \int\!\!\rmn{d}M\frac{\rmn{d}n}{\rmn{d}M}b(M) \left(\frac{M}{\bar\rho_\rmn{m}}\right)^2
\left|\tilde{u}_M(k;\chi)\right|^2,
\label{eq:ps_bw}
\end{equation}
where $\bar\rho_\rmn{m}$ is the comoving mass density and
$\tilde{u}_M(k;\chi)$ is the Fourier-transformed counterpart of the
normalized Navarro-Frenk-White \citep[NFW;][]{Navarroetal:97} profile 
for halos of mass $M$ and at redshift $z$ ($\chi=\chi(z)$): 
$u_M(r)=\rho_{\rm NFW}(r; M)/M$.
$\rmn{d}n/\rmn{d}M$ is the halo mass function and $b(M)$ is the 
linear halo bias for which we throughout this paper use 
 the fitting formula of \cite{ShethTormen:99}.
$P_{\rmn{m}}^\rmn{L}(k;\chi)$ is the linear matter power spectrum and 
$\tilde{W_\rmn{s}}(k\chi\Theta_{\rm s})$ is the Fourier transform of the
survey window function, for which we simply consider a circle-shaped
survey geometry with a radius of $\Theta_\rmn{s}$;
$\tilde{W_\rmn{s}}(x)= 2J_1(x)/x$.  

The first term of Eq.~(\ref{eq:pscovall}) is the Gaussian covariance
term that vanishes when $l\neq l'$,  i.e. no correlation between the power
spectra of different multipole bins. The second term is a non-Gaussian
term arising from the lensing trispectrum (the 4-point correlation
function) and describes correlations between different multipole bins
\citep{Scoccimarroetal:99}. The third term is another non-Gaussian error
due to the HSV effect, which arises from the mode coupling of the
Fourier mode of interest with super-survey modes comparable with or
beyond the survey region via the halo bias theory
\citep{Satoetal:09,Kayoetal:13} \citep[also see][for the derivation in a
mathematically rigorous manner]{TakadaHu:13}. Although there is
another sample variance arising from super-survey modes with Fourier
modes in the weakly nonlinear regime, 
relevant for angular scales around $l\sim 100$,
the effect
is very small at $l \simgt 1000$
\citep{TakadaJain:09}. Hence we ignore this contribution. As carefully
studied in \cite{Satoetal:09}, the covariance formula
(Eq.~\ref{eq:pscovall}) 
well reproduces
the simulation
results. Note that, if ignoring the HSV contribution, the analytical
model significantly underestimates the covariance amplitudes by up to a
factor of 2--3 compared to the simulation results.

\subsubsection{Bispectrum covariance}

Following the formulation in \cite{Kayoetal:13}, we can derive the
covariance matrix of lensing bispectra for lensing tomography case. 
Similarly to the power spectrum case
(Eq.~\ref{eq:pscovall}), the bispectrum covariance has three
contributions: 
\begin{eqnarray}
\rmn{Cov}[B_{(ijk)}(l_1, l_2, l_3),B_{(i'j'k')}(l_1', l_2', l_3')]
 = \rmn{Cov}_\rmn{Gauss}^\rmn{BS} + \rmn{Cov}_\rmn{NG}^\rmn{BS} +
 \rmn{Cov}_\rmn{HSV}^\rmn{BS}.
 \label{eq:covbiall}
\end{eqnarray}
In the following we give the expression of each term.

The first term of Eq.~(\ref{eq:covbiall}) is the contribution arising
from products of the lensing power spectra, which we call the Gaussian
error contribution:
\begin{eqnarray}
\rmn{Cov}_\rmn{Gauss}^\rmn{BS}
 &\equiv&
 \frac{\Omega_\rmn{s}}{N_\rmn{trip}(l_1, l_2, l_3)}
 \left[P^\rmn{obs}_{(ii')}(l_1)\delta^K_{l_1l_1'}
  \left\{
   P^\rmn{obs}_{(jj')}(l_2)P^\rmn{obs}_{(kk')}(l_3)\delta^K_{l_2l_2'}\delta^K_{l_2l_2'}
   +P^\rmn{obs}_{(jk')}(l_2)P^\rmn{obs}_{(kj')}(l_3)\delta^K_{l_2l_3'}\delta^K_{l_3l_2'}
  \right\}
  \right. \nonumber\\
 &&
  \left.
   + \left\{\mbox{2 terms obtained by perm. of } (i'\leftrightarrow j',
      l_1'\leftrightarrow l_2') \right\}+
   \left\{\mbox{2 terms by } (i'\leftrightarrow k',
    l_1'\leftrightarrow l_3') \right\}
  \right],
 \label{eq:covbigauss}
\end{eqnarray}
where $P^{\rm obs}$ includes the shape noise as given by Eq.~(\ref{eq:ps_obs}).
Here, $N_\rmn{trip}(l_1, l_2, l_3)$ is the number of independent
combinations of three vectors ($\bl_1,\bl_2,\bl_3$)
that form 
a given triangle configuration within their bin widths in Fourier
space (again we mean 
 by
`independent' that the triplets are discriminated by the fundamental mode of a
given survey area). For the limit of $l_1,l_2,l_3\gg l_f$,
$N_\rmn{trip}$ is approximated in \citet{Kayoetal:13} as
\begin{equation}
N_\rmn{trip}(l_1,l_2,l_3)
\simeq \frac{\Omega_\rmn{s}^2l_1l_2l_3\Delta l_1 \Delta l_2 \Delta
l_3}{2\upi^3 \sqrt{2l_1^2l_2^2+2l_1^2l_3^2+2l_2^2l_3^2-l_1^4-l_2^4-l_3^4}},
\label{eq:trip}
\end{equation}
where $\Delta l_i$ is the bin width of the $i$-th side length. We here
note that, although some bispectra, for instance
$B_{(112)(l_1,l_2,l_3)}$ and $B_{(121)}(l_1,l_2,l_3)$ ($l_1\ne l_2\ne
l_3$), 
have exactly the
same ensemble-average expectation value as can be found from
Eq.~(\ref{eq:bispectrum}), the above equation (Eq.~\ref{eq:covbigauss})
shows that their covariance elements are different
and therefore the bispectra do
carry different information. 

The second term of Eq.~(\ref{eq:covbiall}) is the non-Gaussian error
contribution arising from terms of $B\times B$ (products of the
bispectra), $P\times T$ and the 6-point correlation function ($P_6$), as
carefully studied in \cite{Kayoetal:13}:
\begin{eqnarray}
\rmn{Cov}_\rmn{NG}^\rmn{BS}
 &\equiv&\frac{2\upi}{\Omega_\rmn{s}}\frac{1}{l_1\Delta l_1}
 \left[
  B_{(i'jk)}(l_1',l_2,l_3)B_{(ij'k')}(l_1,l_2',l_3')\delta^K_{l_1l_1'}
  + B_{(j'jk)}(l_2',l_2,l_3)B_{(i'ik')}(l_1',l_1,l_3')\delta^K_{l_1l_2'}
  \right.
  \nonumber \\
 &&
\hspace{2em}
\left.  + B_{(k'jk)}(l_3',l_2,l_3)B_{(i'j'i)}(l_1',l_2',l_1)\delta^K_{l_1l_3'}
+\{\mbox{3 terms by } (i\leftrightarrow j,
  l_1\leftrightarrow l_2)\} +\{\mbox{3 terms by } (i\leftrightarrow k,
  l_1\leftrightarrow l_3)\}\right]\nonumber \\
 && +\frac{2\upi}{\Omega_\rmn{s}}\frac{1}{l_1\Delta l_1} 
  \left[P^\rmn{obs}_{(ii')}(l_1)T_{(jkj'k')}(l_2,l_3,l_2',l_3')\delta^K_{l_1l_1'}
   +P^\rmn{obs}_{(ij')}(l_1)T_{(jki'k')}(l_2,l_3,l_1',l_3')\delta^K_{l_1l_2'}
  \right.\nonumber\\
 && 
\hspace{2em}
  +P^\rmn{obs}_{(ik')}(l_1)T_{(jki'j')}(l_2,l_3,l_1',l_2')\delta^K_{l_1l_3'}
\left.
 +\{\mbox{3 terms by } (i\leftrightarrow j,
  l_1\leftrightarrow l_2)\} +\{\mbox{3 terms by } (i\leftrightarrow k,
  l_1\leftrightarrow l_3)\}\right]\nonumber \\
&&+\frac{1}{\Omega_\rmn{s}}\int\!\!\frac{\rmn{d}\psi}{2\upi}P_{6(ijki'j'k')}(l_1,
 l_2, l_3, l_1', l_2', l_3'; \psi).
\label{eq:covbing}
\end{eqnarray}
See Fig.~1 in \cite{Kayoetal:13} for the physical interpretation of
each term. Note that $\psi$ is the angle between the two triangle
configurations of the two bispectra. The above equation shows that 
similar bispectra such as 
$B_{(112)}(l_1,l_2,l_3)$ and
$B_{(121)}(l_1,l_2,l_3)$ have different covariance elements as in 
 the Gaussian covariance
elements.

The third term of Eq.~(\ref{eq:covbiall}) is the HSV contribution to the
bispectrum covariance:
\begin{eqnarray}
\rmn{Cov}_\rmn{HSV}^\rmn{BS}
 &\equiv&
\int\!\!\rmn{d}\chi
W_{(i)}W_{(j)}W_{(k)}W_{(i')}W_{(j')}W_{(k')}\chi^{-8}
\nonumber\\
&&\times
B_{\rmn{m}}^{\rm 1hb}(l_1/\chi, l_2/\chi, l_3/\chi;\chi)
B_{\rmn{m}}^{\rm 1hb}(l'_1/\chi, l'_2/\chi, l'_3/\chi;\chi)
  \int\!\!\frac{k\rmn{d}k}{2\upi}P_{\rmn{m}}^\rmn{L}(k;\chi)\left|
\tilde{W_\rmn{s}}(k\chi\Theta_{\rm s})
\right|^2,
\label{eq:covbihsv}
\end{eqnarray}
where we have defined the notation $B_{\rmn{m}}^{\rm 1hb}$ to denote the 1-halo
term of matter bispectrum, weighted by the halo bias, similarly to
Eq.~(\ref{eq:ps_bw}):
\begin{equation}
B_{\rmn{m}}^{\rm 1hb}(k_1, k_2, k_3;\chi)\equiv \int\!\!\rmn{d}M\frac{\rmn{d}n}{\rmn{d}M}b(M) \left(\frac{M}{\bar\rho_\rmn{m}}\right)^3
\tilde{u}_M(k_1;\chi)\tilde{u}_M(k_2;\chi)\tilde{u}_M(k_3;\chi).
\end{equation}
For the HSV covariance, the similar bispectra such as 
$B_{(112)}(l_1,l_2,l_3)$ and 
$B_{(121)}(l_1,l_2,l_3)$ have the exactly same HSV terms, and therefore
the bispectra are highly correlated with each other.

Using Eqs.~(\ref{eq:covbigauss}), (\ref{eq:covbing}) and
(\ref{eq:covbihsv}), we can compute the covariance matrix elements of
lensing bispectra, which we will use for the following
results. \citet{TakadaJain:04} considered only the Gaussian covariance
(Eq.~\ref{eq:covbigauss}), and we will study
how including the non-Gaussian errors degrades parameter forecasts.

\subsubsection{Cross-covariance between power spectrum and bispectrum}

The lensing power spectrum and bispectrum are not totally independent,
as they arise from the same large-scale structure. Hence we need to
properly take
account of their cross-covariance.

Extending the formulation in \cite{Kayoetal:13}, we can similarly derive
the cross-covariance when including lensing tomography:
\begin{equation}
 \rmn{Cov}\left[P_{(ij)}(l), B_{(i'j'k')}(l_1, l_2, l_3)\right]
  = \rmn{Cov}_\rmn{NG}^\rmn{P-B}+\rmn{Cov}_\rmn{HSV}^\rmn{P-B}.
\label{eq:ps-bispcov}
\end{equation}
The first term is the non-Gaussian error term arising from terms of 
$P\times B$ and the 5-point correlation function $P_5$: 
\begin{eqnarray}
\rmn{Cov}_\rmn{NG}^\rmn{P-B}
 &\equiv&
 \frac{2\upi}{\Omega_\rmn{s}}\frac{1}{l_1\Delta l_1}
 \left[
  P^\rmn{obs}_{(i'j)}(l)B_{(ij'k')}(l,l_2,l_3)\delta^K_{ll_1}
  +P^\rmn{obs}_{(ii')}(l)B_{(jj'k')}(l,l_2,l_3)\delta^K_{ll_1} 
\right.\nonumber\\
 &&
  \left.
   +\left\{\mbox{2 terms by } (i'\leftrightarrow j',l_1'\leftrightarrow
     l_2')\right\}
   +\left\{\mbox{2 terms by } (i'\leftrightarrow k',l_1'\leftrightarrow l_3')\right\}
  \right]
+\frac{1}{\Omega_\rmn{s}}\int\!\!\frac{\rmn{d}\psi}{2\upi}P_{5(iji'j'k')}(l,l,l_1,l_2,l_3;\psi).
\label{eq:cross-ng}
\end{eqnarray}
The second term is the HSV contribution defined as
\begin{equation}
\rmn{Cov}_\rmn{HSV}^\rmn{P-B}
 \equiv
\int\!\!\rmn{d}\chi W_{(i)}W_{(j)}W_{(i')}W_{(j')}W_{(k')}\chi^{-6}
P_{\rmn{m}}^{\rm 1hb}(l/\chi;\chi)
B_{\rmn{m}}^{\rm 1hb}(l'_1/\chi, l'_2/\chi, l'_3/\chi;\chi)
\int\!\!\frac{k\rmn{d}k}{2\upi}P_{\rmn{m}}^\rmn{L}(k;\chi)\left|
\tilde{W_\rmn{s}}(k\chi\Theta_{\rm s})
\right|^2.
\label{eq:covps-bihsv}
\end{equation}
Again \citet{TakadaJain:04} did not include the cross-covariance, while
we properly take it into account for parameter forecasts. 

\subsection{Halo model approach}

As described up to the preceding section, the power spectrum and
bispectrum covariance calculations require to compute the four-, five-
and six-point correlation functions of the underlying matter
distribution in addition to the power spectrum and bispectrum. Since
most of the useful information in weak lensing arises from scales
that are affected by nonlinear clustering, theoretical models of the
higher-order matter spectra 
need to be fairly accurate for such nonlinear
scales, up to $k\sim 1~h/{\rm Mpc}$ \citep{HutererTakada:05}. Following
the method in \cite{Kayoetal:13}, we employ the halo model approach
\citep{PeacockSmith:00,Seljak:00,MaFry:00,Scoccimarroetal:01,TakadaJain:03a,TakadaJain:03} to model
the higher-order functions of matter distribution \citep[also see][for a
review]{CooraySheth:02}. In brief, to compute model predictions for the
lensing power spectrum and bispectrum, we employ the full halo model
calculation; we included the one- and two-halo term contributions for
the power spectrum, while we included the one-, two- and three-halo term
contributions for the bispectrum. To compute the four-, five- and
six-point correlation functions in Fourier space for
the covariance calculations, we use only their one-halo term
contributions and ignore the different halo-term contributions, because
the higher-order functions are important only on small angle scales in
the nonlinear regime, where the one-halo term gives a dominant
contribution.  To compute the HSV contribution to the covariances, we
use the third term of Eq.~(\ref{eq:pscovall}) and
Eqs.~(\ref{eq:covbihsv}) and (\ref{eq:covps-bihsv}).

Although the halo model is an empirical method of modeling the nonlinear
clustering, previous work has shown that the halo model predictions
give a 10--20\% level agreement with the simulation results 
in the power spectrum and bispectrum
amplitudes \citep[][also see
references therein]{CoorayHu:01,TakadaJain:03,Takahashietal:12}.  In
addition, \citet{Kayoetal:13} recently studied the bispectrum covariance
using the halo model and showed that the model predictions are again in
good agreement with the simulation results to within 10--20\% level
accuracy in their amplitudes. Hence, we believe that the halo model is
sufficient for our purpose. Since we need to consider many number of
bispectra of different triangles (up to $\sim 10^4$ bispectra in this
paper) to include the full information, the halo model seems a unique,
feasible way in practice 
for the covariance computation; in other words, it requires
too many different realizations to reliably compute the covariance
matrices.

\subsection{Fisher matrix formalism}

\begin{table*}
\begin{center}
 \begin{tabular}{l|l|lll|l|l}\hline\hline
multipole range   & $10\le l \le 1000$
& \multicolumn{3}{c}{$10\le l \le 2000$}
& $10\le l \le 3000$
& $10\le l \le 4000$
\\
redshift bins
&\hspace{1em}
3 $z$-bins 
\hspace{1em}
& \hspace{1em}
1 $z$-bin & 2 $z$-bins &  3 $z$-bins 
\hspace{1em}
& \hspace{1em} 3 $z$-bins \hspace{1em}
& \hspace{1em} 3 $z$-bins \hspace{1em}
\\ \hline
the number of $\bmath{P}(l)$ & 
\hspace{1em}
108  
\hspace{1em}
& 
\hspace{1em}
20 & 60 & 120 
\hspace{1em}
& \hspace{1em}
132 
\hspace{1em}
& 
\hspace{1em}
138
\hspace{1em}
\\
the number of $\bmath{B}(\bmath{l})$ & 
\hspace{1em}
5499 
\hspace{1em}
& 
\hspace{1em}
330 & 2106 & 6527
\hspace{1em}
& \hspace{1em}
7627 
\hspace{1em}
& 
\hspace{1em}
8204
\hspace{1em}
\\ 
\hline \hline
 \end{tabular}
\caption{Number of different power spectra or bispectra 
considered in the parameter forecast, for cases with and without
 lensing tomography and for different maximum multipole $l_{\rm max}$. 
Here we mean by ``different''
that the different power spectra or bispectra have different
covariance matrix elements, and therefore carry complementary information on
 cosmological parameters. We employ the logarithmically-spacing
 multipole bins in the given multipole range, and consider 1 (i.e. no
 tomography), 2 or 3  redshift bins (see text for details). 
As  more tomographic redshift bins are included and  the higher
 $l_{\rm max}$ value is considered, 
the number of different bispectra rapidly
 increases. 
\label{tab:number_ps-bisp}
}
\end{center}
\end{table*}

We use the Fisher information matrix analysis to assess an ability of a
given imaging survey for constraining cosmological parameters from
measurements of power spectrum or bispectrum and their joint
measurement, including tomographic information.  To do this, we include
all available combinations of redshift bins, multipole bins and triangle
configurations in the power spectra and bispectra, taking account of
the non-Gaussian error covariance matrices.

As the lensing observables, we define the following data vector: 
\begin{equation}
\bmath{D}\equiv \left\{\bmath{P},\bmath{B}\right\}
\label{eq:vec_data}
\end{equation}
where $\bmath{P}$ is the data vector containing the power spectra of
different multipole bins and redshift bins and $\bmath{B}$ is similarly
the vector containing the bispectra of different triangle configurations
and redshift bin combinations. When we consider $N$ multipole bins over
a range of $l_1\le l\le l_{\rm max}$ and $n_s$ tomographic redshift
bins ($i=1,\cdots, n_s$), the data vectors are given as
\begin{eqnarray}
\bmath{P}&\equiv& \left\{
P_{(11)}(l_1), P_{(12)}(l_1), \cdots, P_{(1n_s)}(l_1),\cdots, 
P_{(n_sn_s)}(l_1), \cdots, P_{(n_sn_s)}(l_{N})
\right\},\nonumber\\
\bmath{B}&\equiv & \left\{
B_{(111)}(l_1,l_1,l_1), 
B_{(112)}(l_1,l_1,l_1), \cdots, 
B_{(11n_s)}(l_1,l_1,l_1), \cdots, 
B_{(122)}(l_1,l_1,l_1), \cdots, 
B_{(n_sn_sn_s)}(l_N,l_N,l_N) 
\right\}.
\end{eqnarray}
Table~\ref{tab:number_ps-bisp} shows how many different power spectra or
bispectra to consider for the parameter forecasts for different $l_{\rm
max}$ values and the case with and without lensing tomography. 
In this
paper, we mainly consider the multipole range of 
$10\le l\le 2000$, beyond which structure formation is more affected by
physics in too deeply nonlinear regime such as baryonic physics
\citep{HutererTakada:05}. We employ logarithmically-spacing multipole
bins, 20 bins in the range $10\le l \le 2000$, and this binning is
sufficient to capture the shape of lensing power spectrum and the
bispectra of different triangle 
configurations. In other words, we have checked that, even if
we employ a finer multipole binning (e.g., double the number of multipole
bins), 
the parameter forecasts are almost
unchanged. The table shows that, as we include tomographic redshift bins
and increase the maximum multipole $l_{\rm max}$, the number of
different bispectra rapidly increase. For 3 redshift-bin tomography
and $l_{\rm max}=2000$, we consider about 6500 bispectra. 
Thus the 1000 ray-tracing
simulation realizations, which we used in our previous work
\citep{Kayoetal:13}, are not sufficient to estimate the covariance
matrix (see below Eq.~\ref{eq:bispectrum} for the way of counting the
different power spectra and bispectra).

The covariance matrix for the data vector is given as 
\begin{equation}
 \bmath{C}^{\rm PS+Bisp}\equiv \left
\langle \bmath{D}\bmath{D}^t
\right\rangle
-\left
\langle \bmath{D}\right\rangle
\left\langle
\bmath{D}^t
\right\rangle
=
\left(
\begin{array}{cc}
\bmath{C}^{\rm PS} & \bmath{C}^{\rm PS-Bisp}\\
\bmath{C}^{\rm PS-Bisp} & \bmath{C}^{\rm Bisp}
\end{array}
\right),
\end{equation}
where the superscript notation ``${}^t$'' denotes its transposed
matrix, $\bmath{C}^{\rm PS}$ and $\bmath{C}^{\rm Bisp}$ are the
covariance matrices of power spectra and bispectra
(Eqs.~\ref{eq:pscovall} and \ref{eq:covbiall}), and 
$\bmath{C}^{\rm PS-Bisp}$ is their cross-covariance matrix 
(Eq.~\ref{eq:ps-bispcov}).

The Fisher information matrix for the joint lensing power spectrum and
bispectrum tomography is defined as
\begin{eqnarray}
 F_{\alpha \beta}^{\rm WL}\equiv 
\frac{\partial \bmath{D}^t}{\partial p_\alpha}
\left[
\bmath{C}^{\rm PS+Bisp}
\right]^{-1}
\frac{\partial \bmath{D}}{\partial p_\beta},
\end{eqnarray}
where $p_\alpha$ is a set of model parameters (cosmological parameters
plus nuisance parameters if included). The above equation involves
products of the data vector and the covariance matrix, and the product
includes summation over different power spectra and bispectra.
The partial
derivatives in the above equation are done by slightly varying each
parameter from the fiducial value, with fixing other parameters to their
fiducial values. The marginalized $1\sigma$ error on the $\alpha$-th
parameter is given as $\sigma^2(p_\alpha)=[(\bmath{F}^{\rm
WL})^{-1}]_{\alpha\alpha}$, where $(\bmath{F}^{\rm WL})^{-1}$ is the
inverse of the Fisher matrix. When considering confidence regions in a
two-parameter subspace, including marginalization over other parameters,
we follow the method described in Section 4.1 in \citet{TakadaJain:04}.

Weak lensing alone cannot constrain all the cosmological parameters
simultaneously due to parameter degeneracies. Hence, as in done in
\cite{OguriTakada:11}, we also include the CMB information expected for
the Planck experiment. We employ the same method in
\citet{OguriTakada:11} to compute the Fisher matrix of the Planck
expected CMB information.
The Fisher matrix for the joint experiment combining the lensing
information and the CMB information is simply given as
$\bmath{F}=\bmath{F}^{\rm WL}+\bmath{F}^{\rm CMB}$. The Thomson
scattering depth $\tau$ is marginalized over before the CMB Fisher
matrix is added to other constraints.

\section{Results}
\label{sec:results}

\subsection{Parameters}
\label{sec:fisher}

As we stated above, we use the Fisher information matrix analysis to
assess an ability of a hypothetical weak lens survey for estimating
cosmological parameters.  The parameter forecast is sensitive to a
choice of parameters to be included as well as to the fiducial model.
We include a fairly wide range of cosmological models that are given by
a set of eight cosmological parameters: 
the density parameters of matter, baryon and dark energy are
$\omh=0.134$, $\obh=0.0226$, and $\ode=0.734$; the dark energy equation
of state is parametrized by $w(z)=w_0+w_a(1-z)$ with their fiducial
values $w_0=-1.0$ and $w_a=0$;
the Hubble constant
$H_0\equiv100h=71.0$~km/s/Mpc.
We model the linear matter power spectrum
following \citet{Takadaetal:06} as
\begin{equation}
 \frac{k^3}{2\upi^2}P_{\rm
  m}^\rmn{L}(k;a)=\delta_\zeta^2\left(\frac{2k^2}{5H_0^2\om}\right)^2[T(k)D(a)]^2
  \left(\frac{k}{k_0}\right)^{n_s-1+(1/2)\alpha_s\ln(k/k_0)}, \label{eq:plinear}
\end{equation}
where $n_s(=0.963)$ is the spectral tilt, $\alpha(=0)$ is the spectral
running index, and $\delta_\zeta(=4.89\times
10^{-5})$ is the normalization parameter for 
the primordial curvature perturbations
 (the number in the
parenthesis is the fiducial value). The primordial power spectrum is
given at the pivot scale $k_0=0.002$~Mpc$^{-1}$ following the WMAP
convention \citep{Komatsuetal:11}. 
$T(k)$ is the transfer function,
$D(a)$ is the linear growth rate, and the functions can be computed
without ambiguity once a cosmological model is specified. 
We use the publicly-available code, Code for Anisotropies in the
Microwave Background \citep[CAMB;][]{Lewisetal:00},
to compute the transfer function of total matter perturbation.
Our fiducial model gives $\sigma_8\simeq 0.80$, which is the present-day
rms of the linear mass fluctuations in a sphere of radius 8$h^{-1}$Mpc.
To compute the Planck CMB Fisher matrix, 
we further include the optical depth parameter $\tau(=0.089)$. 

As for a galaxy survey, we employ survey parameters that resemble
the planned weak lens surveys, the Subaru HSC Survey and the Euclid
survey. For the Subaru HSC survey, we employ
$\Omega_\rmn{s}=1500$~deg$^2$, $\bar{n}_g=20$~arcmin$^{-2}$, and
$\langle z \rangle=1$ for the survey area and the mean number density
and mean redshift (depth) of galaxies (Eq.~\ref{eq:nz}), respectively.
When considering lensing tomography of $n_s$ redshift bins, we divide the
galaxy redshift distribution in such a way that each redshift bin has
equal number density given by $\bar{n}_g/n_s$. For 
the Euclid survey, we assume $\Omega_{\rm s}=15000$~deg$^2$,
$\bar{n}_g=10$~arcmin$^{-2}$ and $\langle z\rangle=0.7$, respectively. 

\subsection{Parameter forecasts}
\label{sec:forecasts}

\begin{table*}
\begin{center}
\begin{tabular}{l|lll c lll c lll} \hline\hline
& \multicolumn{3}{c}{1 $z$-bin (no tomography)}
& \hspace{1em}
& \multicolumn{3}{c}{2 $z$-bins}
& \hspace{1em}
& \multicolumn{3}{c}{3 $z$-bins} \\ 
parameter & PS & Bisp & PS+Bisp
&
& PS & Bisp & PS+Bisp
&
& PS & Bisp & PS+Bisp 
\\ \hline
$\sigma(\Omega_{\rm de})$ 
& 0.083 & 0.086 & 0.071 (14\%) 
&& 0.043 & 0.050 & 0.036 (16\%)
&& 0.036 & 0.042 & 0.032 (11\%)\\
$\sigma(w_{\rm pivot})$   
& 0.26  & 0.29  & 0.14  (46\%)
&& 0.066 & 0.086 & 0.052 (21\%)
&& 0.060 & 0.080 & 0.048 (20\%)\\
$\sigma(w_0)$ 
& 0.59  & 0.64  & 0.55  (7\%)
&& 0.51  & 0.62  & 0.38 (25\%)
&& 0.38 & 0.52 & 0.32 (16\%)\\
$\sigma(w_a)$  
& 1.4   & 1.8   & 1.1  (20\%)
&& 1.3   & 1.6   & 0.94 (28\%)
&& 0.96 & 1.4 & 0.78 (19\%)\\
FoM                       
& 2.7   & 1.9   & 6.4   (137\%)
&& 11    & 7.2   & 20   (82\%)
&& 17 & 9.2 & 27 (59\%)\\ 
\hline \hline
\end{tabular}
\caption{Summary of marginalized errors on dark energy parameters,
 expected for the power spectrum (PS), the bispectrum (Bisp) and the
 joint measurement (PS+Bisp) with and without lensing tomographic
 information, for a Subaru HSC-type survey that is characterized by
 $\Omega_{\rm s}=1500~$deg$^2$, $\bar{n}_g=20~$arcmin$^{-2}$, and
 $\langle z_s\rangle=1$. Here we consider one redshift bin (no tomography
 case) and 2 and 3 redshift bins for the lensing tomography.  The
 1$\sigma$ error includes marginalization over other parameters. The
 number in the bracket for PS+Bisp error is the fractional improvement
 compared to the error from the power spectrum information alone (PS),
 i.e. quantifying the complementarity of the lensing bispectrum.
 \label{tab:paras}}
\end{center}
\end{table*}

\begin{table*}
\begin{center}
\begin{tabular}{l|lll  lll c lll  lll} \hline\hline
& \multicolumn{6}{c}{1 $z$-bin (no tomography)}
& \hspace{1em}
& \multicolumn{6}{c}{3 $z$-bins}\\
& \multicolumn{3}{c}{HSC}
& \multicolumn{3}{c}{Euclid}
&
& \multicolumn{3}{c}{HSC}
& \multicolumn{3}{c}{Euclid} \\
parameter & PS & Bisp & PS+Bisp
& PS & Bisp & PS+Bisp
&
& PS & Bisp & PS+Bisp 
& PS & Bisp & PS+Bisp 
\\ \hline
$\sigma(\Omega_{\rm de})$ 
& 0.083 & 0.086 & 0.071 (14\%) 
& 0.046 & 0.046 & 0.038 (17\%)
&& 0.036 & 0.042 & 0.032 (11\%)
& 0.020 & 0.023 & 0.015 (25\%)\\
$\sigma(w_{\rm pivot})$ 
& 0.26  & 0.29  & 0.14  (46\%)
& 0.13 & 0.14 & 0.069 (47\%)
&& 0.060 & 0.080 & 0.048 (20\%)
& 0.035 & 0.044 & 0.028 (20\%)\\
$\sigma(w_0)$ 
& 0.59  & 0.64  & 0.55  (7\%)
& 0.41 & 0.52 & 0.32 (22\%)
&& 0.38 & 0.52 & 0.32 (16\%)
& 0.26 & 0.37 & 0.16 (38\%)\\
$\sigma(w_a)$ 
& 1.4   & 1.8   & 1.1  (20\%)
& 1.0 & 1.5 & 0.68 (32\%)
&& 0.96 & 1.4 & 0.78 (19\%)
& 0.70 & 1.0 & 0.41 (41\%)\\
FoM
& 2.7   & 1.9   & 6.4  (137\%)
& 7.9 & 4.7 & 21 (166\%)
&& 17 & 9.2 & 27 (59\%)
& 41 & 22 & 88 (115\%)\\
\hline \hline
\end{tabular}
\caption{Similar to the previous table (Table~\ref{tab:paras}), but
 shows the comparison of parameter forecasts for the HSC Survey and the
 Euclid Survey, where the Euclid Survey is characterized by $\Omega_{\rm
 s}=15000~$deg$^2$, $\bar{n}_g=10$~arcmin$^{-2}$ and $\langle
 z\rangle=0.7$. 
\label{tab:paras_hsc-vs-euclid}
}
\end{center}
\end{table*}

\begin{figure}
\centering \includegraphics[width=0.9\textwidth]{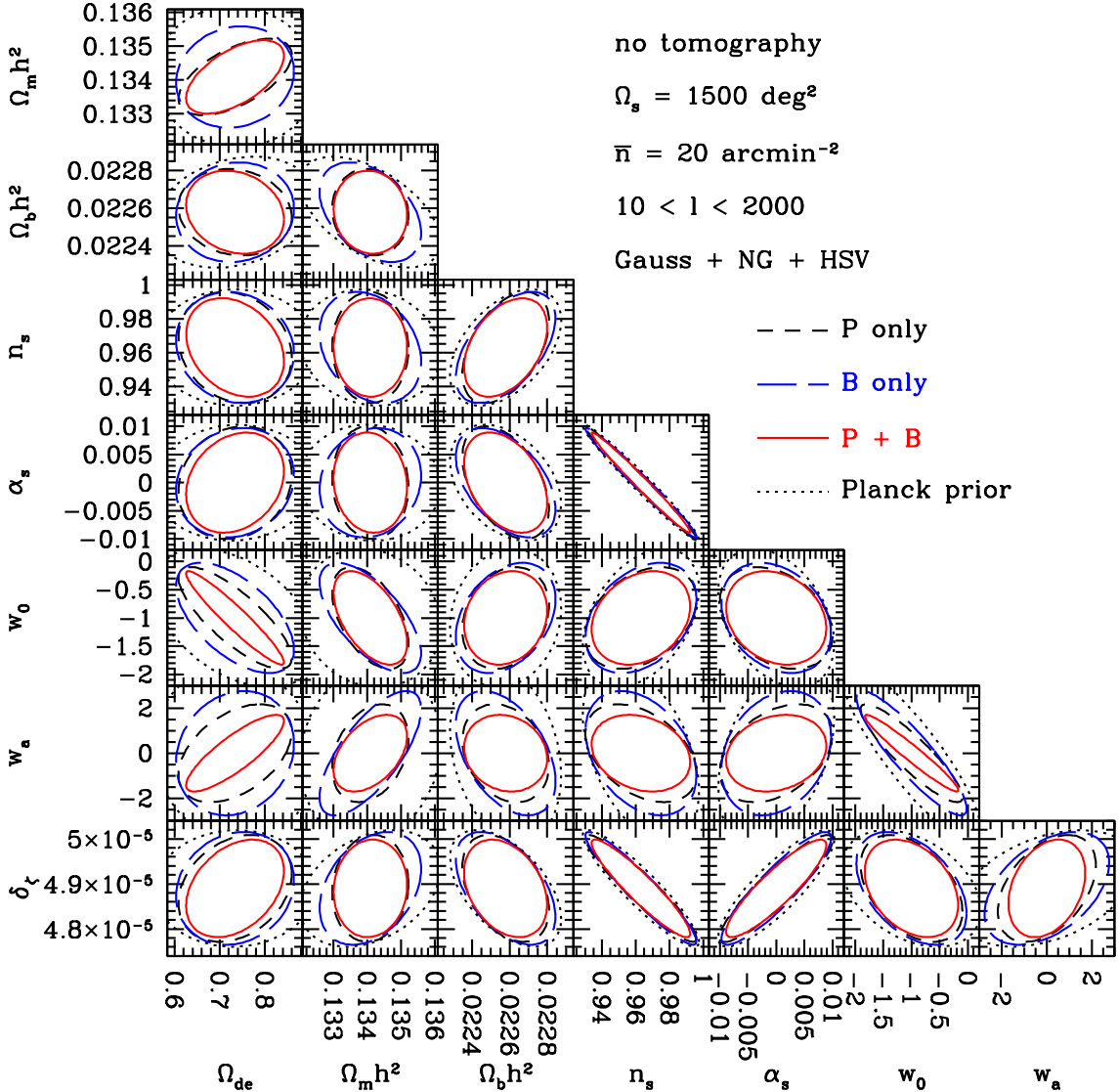}
\caption{Fisher-forecasted error ellipses in each two-parameter
subspace, marginalized over other parameters. The dotted contours are
for the Planck-type CMB information alone. The other contours are the
results expected when combining the CMB information with either of the
lensing power spectrum (dashed, labelled as ``P only'') or the bispectrum
(long-dashed, ``B only'') alone or the joint measurement (solid,
``P+B''), where we employed $\Omega_{\rm s}=1500~$deg$^2$,
$\bar{n}_g=20~$arcmin$^{-2}$, and $\langle z\rangle=1$ for survey
parameters of the hypothetical Subaru HSC survey.  We include the
lensing information over the range of multipole, $10\le l \le 2000$, and
do not include lensing tomography, i.e. consider one redshift bin of
source galaxies.  } \label{fig:all_1bin}
\end{figure}

Fig.~\ref{fig:all_1bin} shows how the lensing power spectrum and
bispectrum information lift parameter degeneracies in the Planck CMB
information for the hypothetical Subaru HSC survey.  For this plot, 
we did not include lensing tomography information, i.e. 1 redshift
bin.  Adding the lensing bispectrum
information to the power spectrum tightens the error ellipses in each
two-parameter sub-space, meaning
that the lensing bispectrum does carry complementary information on
cosmological parameters to the power spectrum. The lensing information
leads to a significant improvement in the dark energy parameters ($\Omega_{\rm
de}, w_0, w_a$), because the parameters are sensitive to the growth of
structure formation from the CMB redshift to low redshifts, while the
primary CMB information cannot well constrain the parameters. 
The other parameters such as the primordial power spectrum parameters are well
constrained by the CMB information. 

\begin{figure}
\centering
\includegraphics[width=0.45\textwidth]{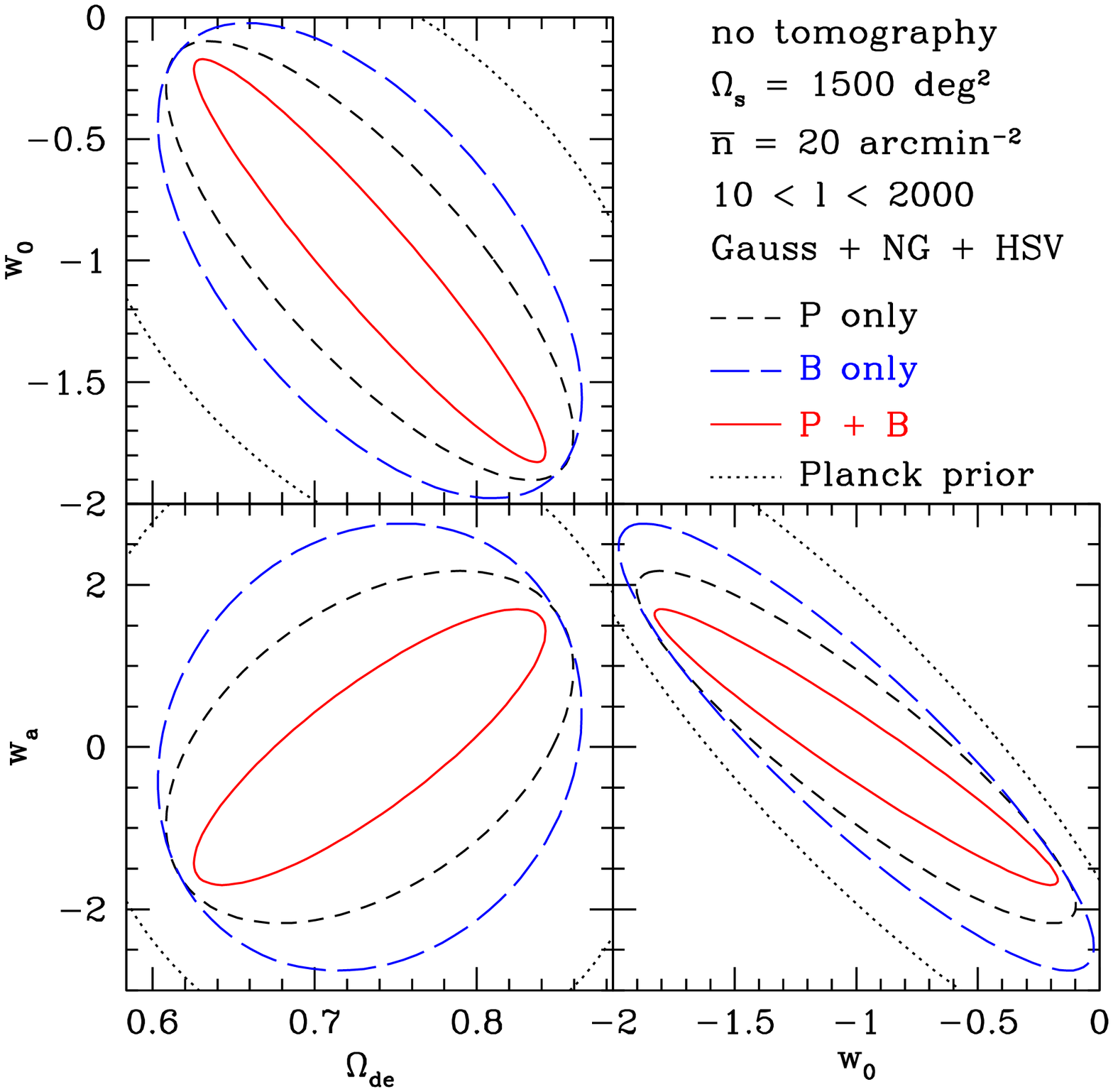}
\includegraphics[width=0.45\textwidth]{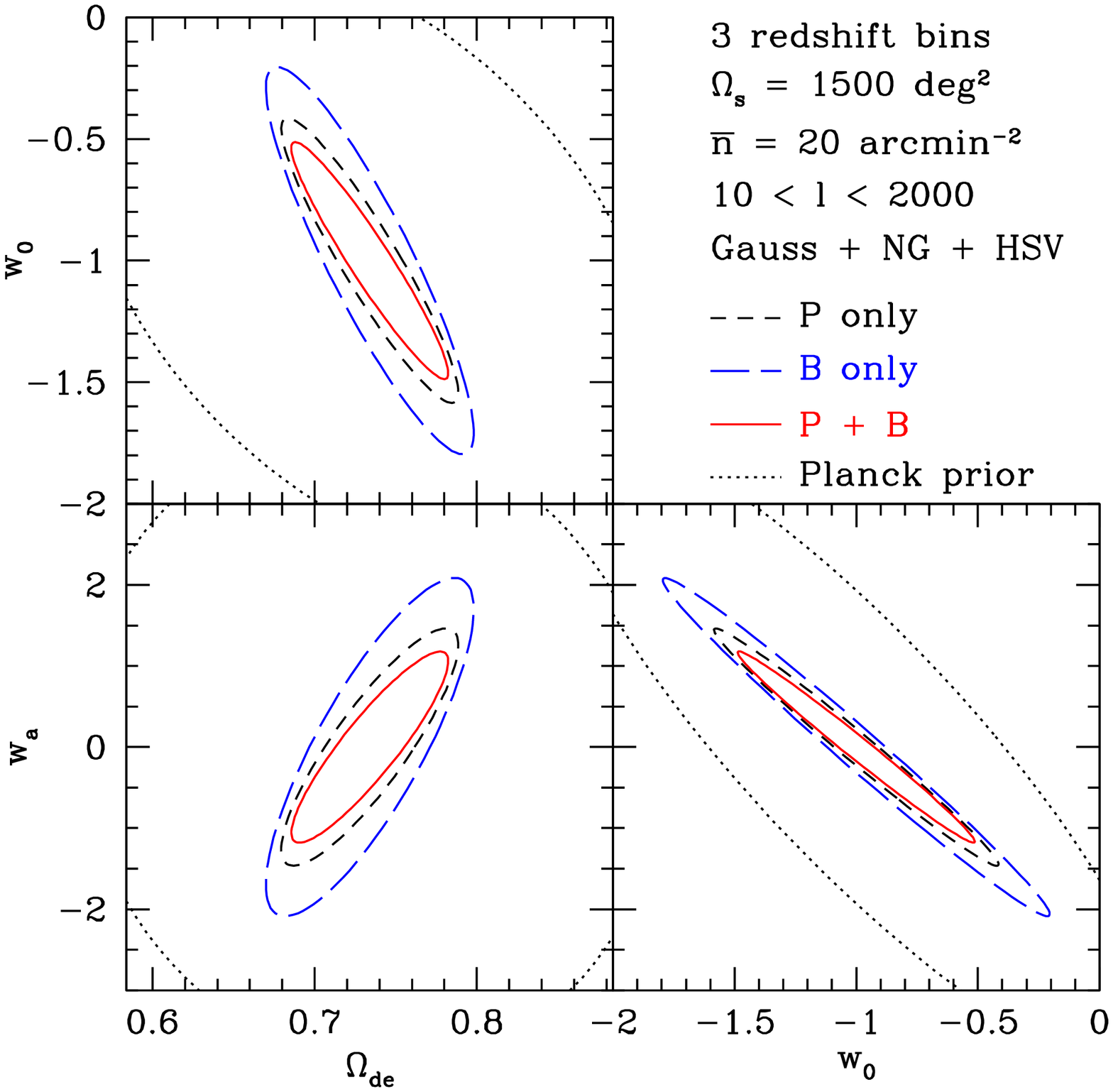}
\caption{{\em Left panel}: Similar to the previous plot, but only for
dark energy parameters ($\Omega_{\rm de}, w_0, w_a$). 
{\em Right panel}: Similar plots, but with lensing
tomography of three redshift bins. 
} \label{fig:deparas}
\end{figure}

In Table~\ref{tab:paras} we show how lensing tomography improves
constraints on dark energy parameters around the fiducial
model (the cosmological constant model). The parameter constraints are
improved by adding tomographic redshift information as well as adding
the bispectrum information. To quantify the accuracy of dark energy
parameters,
we employ the DETF dark energy figure-of-merit (FoM) in \citet{DETF}:
FoM$\equiv 1/[\sigma(w_{\rm pivot})\sigma(w_a)]$. Here
$w_{\rm pivot}$ is the dark energy equation of state parameter at pivot
redshift $z_{\rm pivot}$ and is defined in such a way that the errors in
$w_{\rm pivot}$ and $w_a$ for given observables are uncorrelated.
The error in $w_{\rm pivot}$ can be computed from the sub-matrix of the
inverted Fisher matrix that contains only the elements of $w_0$ and
$w_a$ \citep[see][for the definition]{HuJain:04}.
Table~\ref{tab:paras} shows that the three redshift-bin
tomography allows for FoM$\simeq 30$ when combining the lensing power
spectrum and bispectrum information, a factor of 10 or 5 improvement compared
to FoM$\simeq 2.7 $ or $6.4$ for the power spectrum alone or the joint
measurement without tomography information, 
respectively. 
Comparing the results for the power spectrum
alone and the joint measurement shows that
adding the bispectrum tomography yields about 60--80\% improvement for
the 2 and 3 redshift-bin tomography cases. Since the dark
energy FoM roughly scales with survey area as FoM$\propto \Omega_{\rm
s}$, the improvement is equivalent to 1.6--1.8 larger survey area if
using the power spectrum information alone. 
Fig.~\ref{fig:deparas} visualizes the improvement of constraints on the
dark energy parameters by adding the tomographic bins and the bispectrum
information; considering multiple tomographic bins greatly
improves the constrains, while the relative impact of adding the
bispectrum becomes less prominent.

\begin{figure}
\centering
\includegraphics[width=0.45\textwidth]{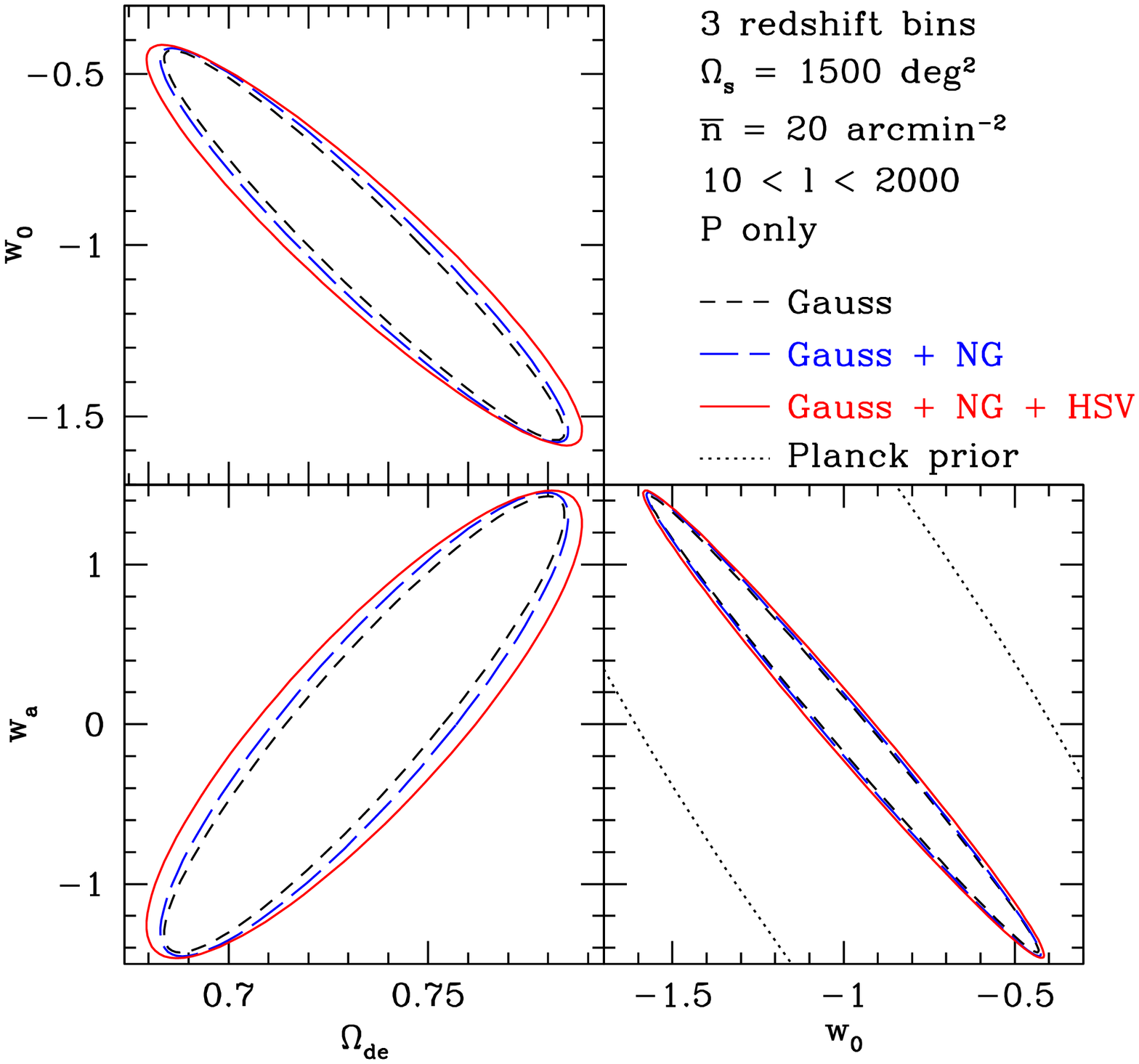}
\includegraphics[width=0.45\textwidth]{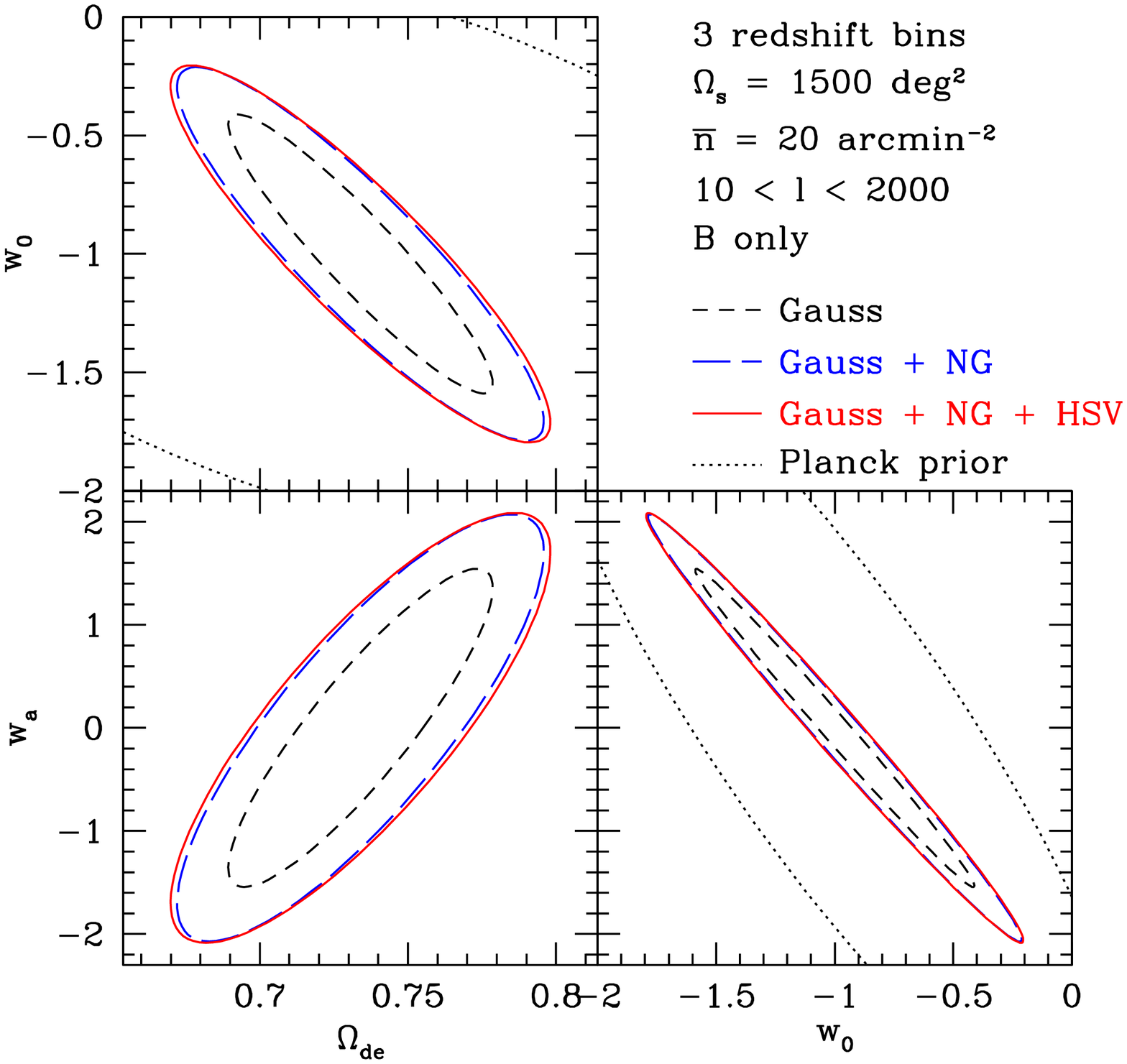}
\includegraphics[width=0.45\textwidth]{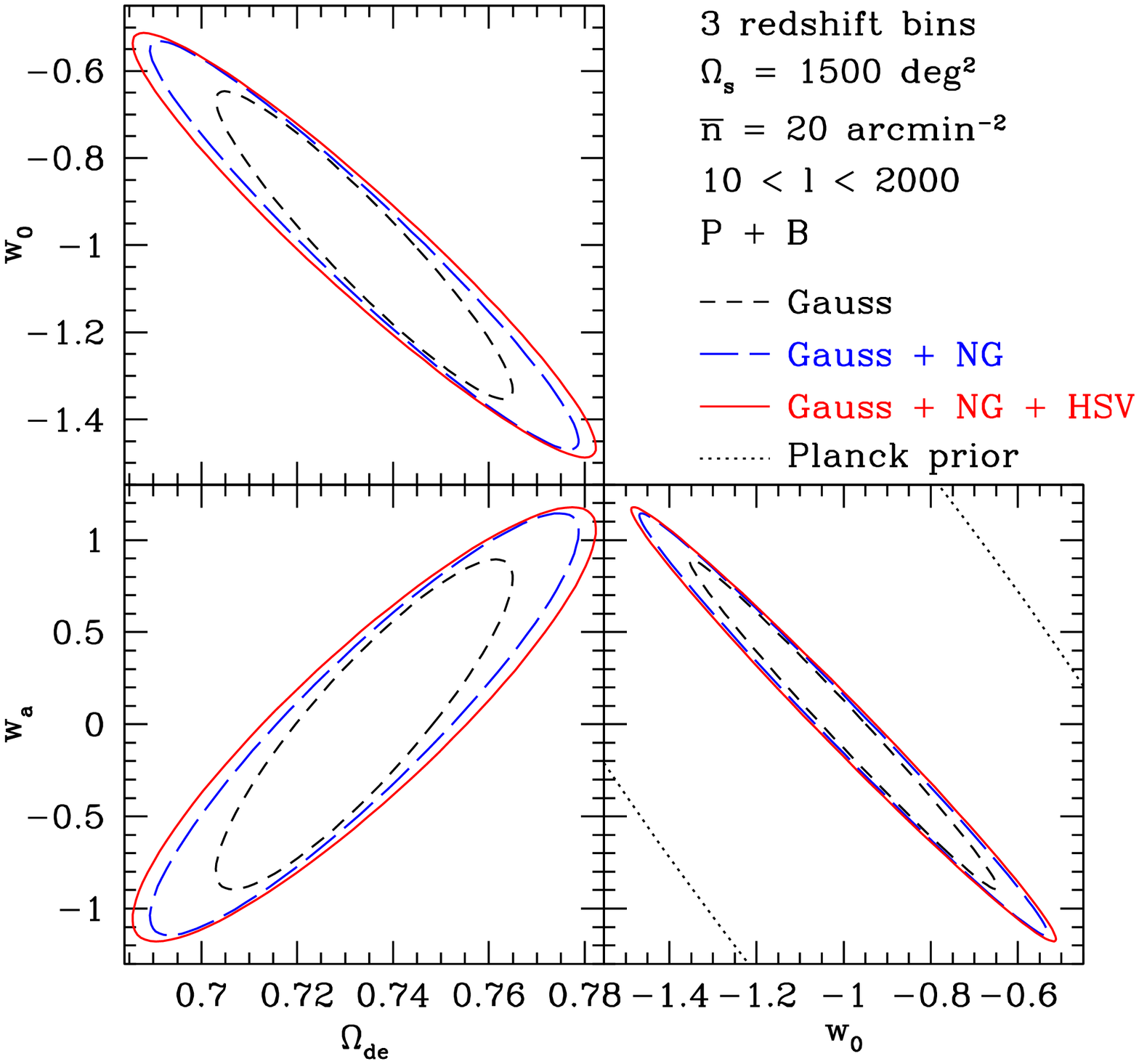}
\includegraphics[width=0.45\textwidth]{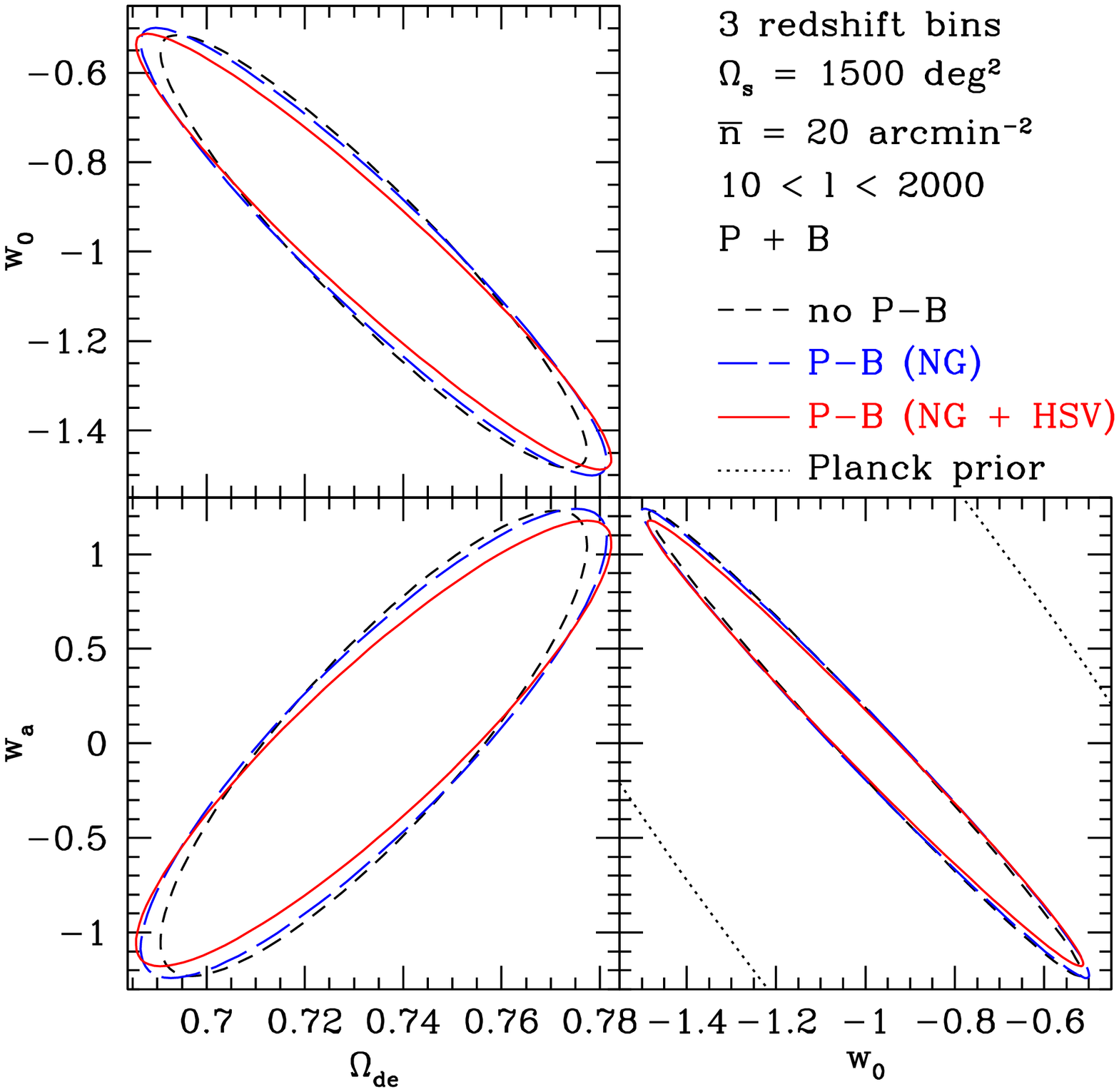}
\caption{The impact of non-Gaussian error covariance on the marginalized
errors of the dark energy parameters. We consider the cases with three
tomographic redshift bins. {\em Upper-left panel}: The result obtained when
 including the power
spectrum tomography alone. 
The solid curves are our fiducial results using the full
covariance matrix of power spectrum, as we have shown in main text,
while the dashed lines show the results obtained by assuming the
 Gaussian covariance matrix, i.e. ignoring the non-Gaussian errors, 
as done in \citet{TakadaJain:04}.  The long-dashed curves are
with non-Gaussian covariance, but ignoring the HSV contribution.  {\em
Upper-right and lower-left panels}: Similar plots, but for the bispectrum
and the joint measurement, respectively.  {\em Lower-right panel}:
The impact of the cross-covariance matrix between the power spectrum and
bispectrum. We include the full covariance elements for the power
 spectrum and the bispectrum, but take different treatments for the
 cross-covariance matrix calculation. 
The solid contours are our fiducial results,
 i.e. including the full covariance. The short-dashed contours, labelled
 ``no P-B'', are the results when ignoring the cross-covariance,
 i.e. no-correlation between the power spectrum and bispectrum. 
The dashed contours, labelled as
 ``P-B (NG)'', are the results when including the cross-covariance
 matrix, but ignoring the HSV contribution.
}
 \label{fig:de_3bin_NG}
\end{figure}
In Fig.~\ref{fig:de_3bin_NG} we study the impact of non-Gaussian errors
on the dark energy parameters. The figure shows how the error ellipses
change if we ignore the non-Gaussian errors (i.e. assuming the
Gaussian error covariances), the HSV
contribution to the non-Gaussian errors, or the cross-covariance between
the power spectrum and the bispectrum. Since the lensing power spectrum
and bispectrum arise from the same large-scale structure in the
light-cone volume, the two are highly correlated with each
other. Comparing the top-left and -right panels reveals that the
non-Gaussian errors yield a more significant degradation in the
parameters for the bispectrum tomography than for the power spectrum.
Although the HSV effect causes a significant degradation in the information
content of the power spectrum or the bispectrum as carefully studied in
\citet{Kayoetal:13}, the impact on the parameter errors is modest after
marginalizing over other parameters. To be more precise,
the HSV effect enlarges a volume of the Fisher error ellipse in
8 dimensional parameter space by a factor of 2 compared to the case
of ignoring the HSV effect, while the projected error for a particular parameter
is enlarged only by about 10\% \citep[$\sim $ 2$^{1/8}$; see also][for a similar
discussion]{TakadaJain:09}.

The results in Table~\ref{tab:paras} and Fig.~\ref{fig:deparas} can be
compared with Table~1 and Fig.~7 in \citet{TakadaJain:04}.
Our results show that adding the bispectrum information to the power
spectrum gives a modest improvement, 
10-20\% for each dark energy parameter, while the latter found about
50\% improvement. 
There are two important differences in
between this work and \citet{TakadaJain:04}. First, we have properly
taken account of the non-Gaussian error covariances to perform the
parameter forecasts, while \citet{TakadaJain:04} considered only the Gaussian
covariance and included the information up to $l_{\rm max}=3000$ instead
of 2000. The non-Gaussian error degrades the parameter
errors. To be more explicit, if assuming the Gaussian
covariance alone for 3 redshift-bin tomography, the power spectrum, the
bispectrum and the joint measurement yield the marginalized errors of
$\sigma(w_0)=0.37$, $0.39$ and $0.23$ ($38\%$) and $\sigma(w_a)=0.94$,
1.0, and $0.59$ (37\%), respectively, which can be compared with
Table~\ref{tab:paras}. Second, we employ the primordial
curvature perturbation for the normalization of the linear matter power
spectrum, while \citet{TakadaJain:04} used the $\sigma_8 $
normalization.
When using the $\sigma_8$ normalization instead of
$\delta_\zeta$, the weak lensing information primarily constrains not
only the
dark energy parameters $(\Omega_{\rm de}, w_0, w_a)$, but also
$\sigma_8$; 
that is, the CMB priors are
less important for these parameters, because the parameters are
sensitive to the growth of structure at low redshifts. 
Hence, adding the bispectrum more
efficiently lifts parameter degeneracies in the four parameters,
yielding a relatively greater improvement in each dark energy parameter
compared to the power spectrum alone.  In fact, we checked that our code
fairly well reproduces the results in \cite{TakadaJain:04} if we follow
the setting of \citet{TakadaJain:04}: the $\sigma_8$ normalization, the
Gaussian covariance, $l_{\rm max}=3000$ for the maximum multipole and
the same survey parameters.
This is encouraging because the details of model ingredients differ in
the two studies.
For example, we used the CAMB to compute the transfer function, while
\citet{TakadaJain:04} used the BBKS transfer function \citep{BBKS}. 
In addition, we
used the halo model to compute the nonlinear bispectrum, while
\citet{TakadaJain:04} employed the hyper extended perturbation theory
\citep{ScoccimarroCouchman:01}.
The results in this paper are also consistent with the results in the
early phase of this project, where the calculation was done using 
 the completely different
codes\footnote{See a talk slides in
\url{http://www.iap.fr/activites/colloques_ateliers/colloque_IAP/ColloqueIAP2007/talks/Friday/Takada.ppt}}.
Thus we believe that our results capture the genuine power
of the lensing bispectrum for constraining cosmological parameters
relative to the power spectrum.

However, we should note that 
our parameter forecasts are quite different from the recent
result in \citet{SatoNishimichi:13}. The study fully relied on the
ray-tracing simulations to compute the lensing power spectrum, the
lensing bispectrum and their response to each cosmological parameter.
They showed rather surprising results. For instance, the dark
energy FoM for the lensing bispectrum tomography is greater than that of
the power spectrum tomography, meaning that the lensing bispectrum is
more powerful than does the power spectrum. They also argued that adding
the bispectrum information yields a factor 2--3 improvement in each
cosmological parameter compared to the power spectrum alone. Their
results imply that the four-point function or the trispectrum can be
even more important than the bispectrum and the power spectrum.  We do
not know where the big differences come from, so a further study is
definitely 
needed.

\begin{figure}
 \centering
\includegraphics[width=0.45\textwidth]{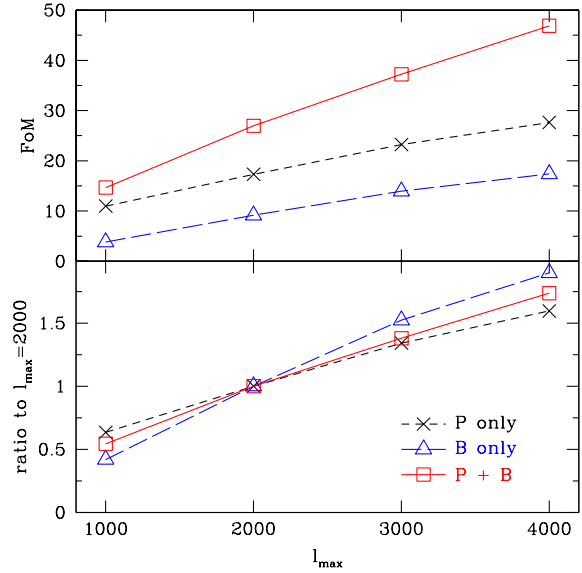}
\caption{
{\em Upper panel}: The dark energy FoM as a function of the maximum
 multipole $l_{\rm max}$, for the power spectrum (P), the bispectrum (B)
 and the joint experiment with 3 redshift-bin tomography, for a Subaru HSC-type
 survey as in Fig.~\ref{fig:deparas}. Note that we 
included the
 information over $10\le l\le l_{\rm max}$. With increasing $l_{\rm
 max}$, adding the bispectrum yields a greater improvement in the FoM
 compared to that of the power spectrum tomography alone. {\em Lower
 panel}: The degradation or improvement relative to the fiducial case
 of 
 $l_{\rm max}=2000$.
\label{fig:lmax}}
\end{figure}

We also consider some cases where the bispectrum tomography increases
its relative importance for the parameter estimation. 
For instance, Table~\ref{tab:paras_hsc-vs-euclid} compares the forecasts
for the HSC- and Euclid-type surveys. Here we assume a lower mean
redshift for the 
Euclid-type survey, $\langle z_s\rangle=0.7$, compared to $\langle
z_s\rangle=1$ for the HSC-type survey.  The weak lensing from a
shallower survey probes large-scale structure at lower redshifts, where
the large-scale structure is more evolved and shows stronger
non-Gaussianity. Hence, the lensing bispectrum becomes more powerful for
a shallower survey.
Another example is in Fig.~\ref{fig:lmax}, which shows how the dark
energy FoM is improved by 
including the power spectrum and bispectrum information up to higher
$l_{\rm max}$, for a Subaru HSC-type survey.  Here we consider the
values of $l_{\rm max}=1000, 2000, 3000$ and
$4000$. The weak lensing field at higher multipoles is
more non-Gaussian, and therefore the lensing bispectrum brings
stronger complementarity to the dark energy parameters.

\section{Conclusion and Discussion}
\label{sec:conclusion}

In this paper, we have extended the halo model based method in
\citet{Kayoetal:13} to model the covariance matrices for the weak
lensing power spectrum and bispectrum when tomographic redshift
information is included. Then we have used the covariance formula to
estimate a genuine power of the lensing bispectra, full the three-point
correlation information, for constraining cosmological parameters when
combined with the power spectrum information, for upcoming weak lensing
surveys. To do this, we included all the bispectrum information built
from all-available combinations of different redshift bins and different
triangle configurations. Thus our study gives an answer on the
best-available complementary information of the three-point correlation
based statistics
of weak lensing relative to the two-point correlation information, the power
spectrum. Any collapsed three-point statistics such as
skewness\footnote{The skewness is a real-space statistics and 
given by the integration of the
bispectrum, weighted by the smoothing function, and carries a partial
information of the different configurations.} or a
partial set of bispectra does not carry as much cosmological information
as what we have shown in this paper.

We have shown that
adding the bispectrum information helps to lift parameter degeneracies
that appear when using the power spectrum information alone, with the
CMB priors (see Figs.~\ref{fig:all_1bin} and \ref{fig:deparas} and
Tables~\ref{tab:paras} and \ref{tab:paras_hsc-vs-euclid}). The parameter
accuracy from the bispectrum information is more degraded by the
non-Gaussian errors than that from the power spectrum (see
Fig.~\ref{fig:de_3bin_NG}). This result would be natural. The weak lensing
power spectrum is primarily sensitive to cosmological parameters and
carries the largest amount of  information on
 the underlying matter distribution,
because large-scale structure originates from the initial Gaussian
field. The bispectrum is a measure of the non-Gaussian features in
late-time large-scale structure that arise from the nonlinear
clustering. In addition, there are too many different bispectra
constructed from the range of multipoles. If assuming the Gaussian error
covariances, it ignores correlations between the bispectra of different
configurations. 
Hence it would be natural that 
the bispectrum is more affected by the non-Gaussian
errors. When adding tomographic redshift information,
improvements in parameter estimation  by adding the bispectrum
information become less significant compared to the improvement without
tomography. Nevertheless, the joint measurement
of the power spectrum and bispectrum for 3 redshift-bin tomography
gives about 60\% improvement in the dark energy FoM compared to the
power spectrum tomography alone, for a Subaru HSC-type survey. This is
promising in a sense that the improvement is equivalent to a factor 1.6
larger survey area if using the power spectrum alone.  However, the
power of lensing bispectrum is not as significant as what was claimed in
\cite{TakadaJain:04}, where a factor 2--3 improvement was found using
the Gaussian covariance. 
Our results imply that even higher-order
functions such as the four-point correlation function bring less
complementary information than does the bispectrum. For a shallower weak
lensing survey that preferentially probes the more evolving large-scale
structure at lower redshift, the bispectrum becomes relatively more
powerful to constrain parameters (see
Table~\ref{tab:paras_hsc-vs-euclid}). For the same reason, when
including the bispectrum up to the higher maximum multipole $l_{\rm
max}$, the bispectrum becomes more useful (Fig.~\ref{fig:lmax}).

The bispectrum depends on cosmological parameters in a different way
from the power spectrum. For instance, the lensing bispectrum is
proportional to the cube of the lensing efficiency function, while the
power spectrum is proportional to the square of the lensing efficiency
function. Further, the lensing bispectrum is more sensitive to
large-scale structure at lower redshifts than is the power spectrum, as
we discussed.  This is the main reason why adding the bispectrum
information to the power spectrum allows for breaking parameter
degeneracies. This complementarity would also be true for systematic
errors inherent in weak lensing measurements such as photometric
redshift errors and imperfect shape measurements; the power spectrum and
bispectrum depend on the systematic errors in different ways. Hence, in
the presence of systematic errors, adding the bispectrum information
would allow for not only improving parameter estimations, but also
calibrate out the systematic errors -- self-calibration from the same
data sets \citep{Hutereretal:06}. The self-calibration issue is worth
exploring, and will be our future study. 

The formulation developed in this paper would have various
applications. It would be interesting to study even higher-order
functions such as the trispectrum and then study how much complementary
information is further added by them, compared to the power
spectrum and bispectrum. The method shown in this paper can be easily
extended to the trispectrum calculations. The abundance of massive halos
are also complementary to the weak lensing information. In particular,
the abundance of massive halos are affected by the same super-survey
modes, through the HSV effect, so combining the weak lensing
correlations with the abundance of massive halos in the same survey
region can be used to correct for the HSV contamination.  As we have
shown, it is very important to take into account the covariance matrices
between the different $n$-point correlations in order to properly count
their independent information. 
Another promising weak
lensing statistics is the cross-correlation of galaxy shapes with
positions of foreground lensing objects (galaxies or clusters) with
known redshifts -- the so-called galaxy-galaxy lensing or cluster-galaxy
lensing \citep{OguriTakada:11,Okabeetal:13}. The cross-correlation can
probe the matter distribution at a particular redshift, the lensing
redshift, i.e. free of the projection effect of large-scale structures
at different redshifts along the line-of-sight direction.  Most previous
work has focused on the two-point correlation functions, but it would
be worth further studying the three-point cross-correlation functions
such as the halo-halo-shear or halo-shear-shear
cross-correlations. Then, by fully taking into account the covariance
matrices between the different cross-correlations, we can estimate the
genuine power of the cross-correlation methods at two- and three-point
level for constraining cosmology. For these studies, the method and
formulation in this paper would be useful. These are our future subjects,
and will be presented elsewhere.

\section*{Acknowledgments}
We thank Bhuvnesh Jain for useful discussion. 
We also thank Masanori Sato for providing us with
the ray-tracing simulation data  used in this work.
MT thanks the Aspen Center for Physics and the NSF Grant \#1066293 for
their warm hospitality, where this work was completed.
This work is supported in part by JSPS KAKENHI (Grant Number: 23340061
and 24740171),  by World Premier International Research Center
Initiative (WPI Initiative), MEXT, Japan, by the FIRST program `Subaru
Measurements of Images and Redshifts (SuMIRe)', CSTP, Japan.  Numerical
computation in this work was partly carried out at the Yukawa Institute
Computer Facility.

\footnotesize{
\bibliographystyle{mn2e}
\bibliography{ms_ref}
}

\end{document}